\documentclass[11pt,a4paper]{article}
\usepackage[sectionbib,authoryear,round]{natbib}
\usepackage{subfig}
\usepackage{color}
\usepackage{graphicx}
\usepackage{amsmath}

\title{Evaluation of spatial audio reproduction schemes for application in hearing aid research}

\date{December 18, 2014}

\definecolor{gg}{rgb}{0.9804,0.9297,0.08594}

\begin{document}


\author{Giso Grimm$^{a,b}$, Stephan Ewert$^a$ and Volker Hohmann$^{a,b}$\\
$^a$ Medizinische Physik, Universit\"at Oldenburg,\\
and Cluster of Excellence ``Hearing4all''\\
$^b$ H\"orTech gGmbH, Oldenburg}

\maketitle

\begin{abstract}
Loudspeaker-based spatial audio reproduction schemes are increasingly
used for evaluating hearing aids in complex acoustic conditions. To
further establish the feasibility of this approach, this study
investigated the interaction between spatial resolution of different
reproduction methods and technical and perceptual hearing aid
performance measures using computer simulations. Three spatial audio
reproduction methods -- discrete speakers, vector base amplitude
panning and higher order ambisonics -- were compared in regular
circular loudspeaker arrays with 4 to 72 channels. The influence of
reproduction method and array size on performance measures of
representative multi-microphone hearing aid algorithm classes with
spatially distributed microphones and a representative single channel
noise-reduction algorithm was analyzed. Algorithm classes differed in
their way of analyzing and exploiting spatial properties of the sound
field, requiring different accuracy of sound field
reproduction. Performance measures included beam pattern analysis,
signal-to-noise ratio analysis, perceptual localization prediction,
and quality modeling.  The results show performance differences and
interaction effects between reproduction method and algorithm class
that may be used for guidance when selecting the appropriate method
and number of speakers for specific tasks in hearing aid research.
\end{abstract}
\footnote{The archived file is not the final published version of the
  article ``Grimm, Giso, Ewert, Stephan, and Hohmann, Volker:
  Evaluation of spatial audio reproduction schemes for application in
  hearing aid research'', in ``Acta Acustica united with Acustica'',
  volume 101, 2015, pp. 842-854(13). \copyright 2015 S. Hirzel Verlag/European
  Acoustics Association. The definitive publisher-authenticated
  version is available online at
  http://www.ingentaconnect.com/content/dav/aaua. http://dx.doi.org/10.3813/AAA.918878.  Readers must contact the
  publisher for reprint or permission to use the material in any form.
} 	


\section{Introduction}

Hearing aids are evolving from simple amplifiers to complex systems
that are aware of the spatial configuration and contents of their
acoustic surroundings \citep{Kates2008}. Moreover, the interaction
between hearing aids and users is gaining increasing attention
\citep{Tessendorf2011}.
This development causes an increase in the complexity of hearing aids
and their use which in turn requires improved evaluation methods in
order to demonstrate the properties and benefits of the systems. One
way of achieving this is to perform evaluations in real acoustic
environments; however, this approach is costly and does not provide
completely controllable and reproducible experimental
conditions. Laboratory studies, on the other hand, are efficient and
reproducible but performance of hearing aid algorithms in real
environments and under laboratory conditions often differs
substantially due to a different subject behavior
\citep[e.g.,][]{Smeds2006} or as a consequence of oversimplification
of the acoustic scenarios \citep{Cord2004,Bentler2005}. This motivates
the reproduction of complex acoustic environments in the laboratory
using loudspeaker-based spatial audio reproduction methods to provide
controllable and reproducible realistic experimental conditions for
hearing aid evaluations. Available reproduction methods, however, have
not yet been evaluated systematically in combination with
multi-microphone hearing aid algorithms.

Typical applications of spatial audio reproduction systems are sound
reinforcement for theaters and cinemas \citep{Brandenburg2004}, music
reproduction \citep{Nettingsmeier2010}, audio reproduction for
computer games \citep{Olaiz2009}, room auralization
\citep{Favrot2009,Favrot2010}, and applications in hearing research
\citep{Seeber2010}. Each application has its own requirements
regarding listening area, tolerance to spatial or timbral artifacts,
maximum technical complexity, computational complexity, and
latency. In contrast to music- and media-reproduction systems,
constraints regarding the size of the listening area are comparably
loose in research applications, since typically only a single listener
or a small group of listeners is addressed simultaneously.  The system
layout in theater and cinema applications often uses an asymmetric
distribution of loudspeakers, e.g., 5.1 \citep{ITU51} or 22.2
\citep{Hamasaki2005,Hamasaki2011}, to achieve a higher spatial
resolution in the frontal hemisphere. Applications in hearing research
commonly use horizontal circular layouts with a regular loudspeaker
distribution.  Methods for generation of the loudspeaker signals
include vector base amplitude panning \citep[VBAP;][]{Pulkki1997},
higher order ambisonics \citep[HOA;][]{Daniel2001} and wave field
synthesis \citep[WFS;][]{Berkhout1993,Spors2008}.  Common to all
loudspeaker-based spatial audio reproduction methods is their limited
spatial resolution due to the finite number of speakers involved in
the reproduction. The type, number and spatial distribution of
artifacts related to this limitation differ substantially between
methods.

Theoretical and perceptual limitations of loudspeaker-based spatial
audio reproduction schemes have been studied extensively
\citep[e.g.,][]{Landone1999,Daniel2003,Ahrens2008}.  Many studies
focus primarily on music reproduction
\citep[e.g.,][]{Bates2007a,Guastavino2004}. Other studies measure
interaural time difference (ITD) and interaural level difference (ILD)
as a predictor of perceptual localization performance
\citep{Daniel2001,Carlsson2004,Pulkki2005,Benjamin2010,Bertet2013}.
These studies thoroughly investigated the perceptual properties of the
reproduction methods. Although the physical sound field is not
correctly reproduced, it was shown that the perceptual impression can
be rendered almost perfectly by exploiting properties of the human
(binaural) hearing system.  However, when the reproduced sound is
processed by a hearing aid algorithm with spatially distributed
microphones prior to presenting it to the subject, both physical
characteristics of the reproduced sound field and perceptual aspects
play a role in assessing the reproduction quality. If the sound field
sampled by the spatially distributed microphones physically deviates
from the original sound field, the function of the algorithm might
be hampered, possibly leading to a decreased algorithm performance and
perceptually audible artifacts. To assess this possibly detrimental
effect, knowledge of the details of multi-microphone processing in
hearing aids is required.

\citet{Hamacher2005} provide an overview of state-of-the-art
algorithms applied in hearing aids. They distinguish between five
classes of algorithms: Directional microphones, single channel noise
reduction, multi-band dynamic compression, feedback suppression, and
classification. In their first class they list all algorithms that use
spatially distributed microphones. These include first-order and
higher-order microphone arrays
\citep[e.g.,][]{Widrow2003,Rohdenburg2007}, extended adaptive
algorithms \citep[e.g.,][]{Elko1995,Spriet2005a}, and binaural noise
reduction schemes
\citep[e.g.,][]{Kollmeier1993,Wittkop1997,Wittkop2003}. Whereas the
functioning of these algorithms explicitly depends on the spatial
properties of the sound field in the small area covered by the
microphones, i.e., close to the head of the hearing-aid user, the
other classes like single channel noise reduction and dynamic
compression, depend only implicitly on the spatial properties of the
surrounding, e.g., as a result of head shadowing, and in the sense
that they potentially modify spatial cues sensed by the
listener. Depending on which spatial aspects of the sound field are
exploited by a specific algorithm, its performance may therefore be
affected differently by the specific limitations of a reproduction
method in reproducing the sound field at the head. Thus, several
algorithm classes need to be tested in combination with different
reproduction methods to assess the interaction between algorithm class
and reproduction method.

Only one simulation study by \citet{Oreinos2013} evaluated performance
of two different hearing aid algorithms with spatially distributed
microphones, an adaptive differential microphone and an interaural
coherence-based directional filter, in a sound field reproduced by a
7th order HOA system. Data show that up to a certain frequency the HOA
system has no significant effect on the algorithm performance, and
that the effect is larger for the adaptive directional microphone than
for the interaural coherence-based directional algorithm. However, while
showing the principle feasibility of the approach, only a single
reproduction method and a fixed array size were considered.

This simulation study systematically evaluates the interaction of
reproduction method, array size and different classes of hearing aid
algorithms with spatially distributed microphones.  The effect of
three spatial audio reproduction methods on the performance of three
conceptually different classes of multi-microphone hearing aid
algorithms, a static binaural beamformer with three microphones at
each ear \citep{Rohdenburg2007}, an adaptive directional microphone
with two microphones at one ear \citep{Elko1995} and an interaural
coherence-based binaural noise reduction algorithm with one microphone
at each ear \citep{Grimm2009a} was assessed.  Furthermore, a standard
single-channel noise reduction scheme \citep{Ephraim1984} was included
in the study. Small to medium-sized loudspeaker arrays with 4 to 72
loudspeakers in a horizontal circular configuration were tested,
representing the size range and type of systems commonly used in
experimental hearing research. Loudspeaker signals were generated with
three different methods, which can be interpreted as three different
methods of spatial interpolation: The selection of the nearest speaker
(NSP) to a virtual sound source uses only a single loudspeaker at a
time. With vector base amplitude panning (VBAP) two loudspeakers are
used to interpolate virtual source positions not covered by a
loudspeaker. In higher order ambisonics (HOA) all loudspeakers
contribute to the spatial image. Unless near-field compensation is
applied to HOA, these methods commonly reproduce phantom sources in
the distance of the loudspeaker array, i.e., they do not encode the
curvature of the wave fronts, and distances can only be coded by
loudness, spectral cues caused by air absorption and, in case of
closed rooms, by the direct-to-reverberant ratio of sounds.  Wave
field synthesis (WFS) is able to reproduce the curvature of wave
fronts by synthesizing the whole sound field of a virtual
source. However, WFS differs in its spatial distribution and type of
artifacts, and thus does not directly compare to the other mentioned
reproduction methods. Specifically, for a comparable amount of
artifacts in a single point for a given frequency bandwidth, WFS
requires a much higher number of speakers than VBAP or HOA; thus WFS
was not considered here.
For an assessment of the effects of reproduction systems, all signals
were generated by convolution of the loudspeaker signals with anechoic
binaural head-related impulse responses (HRIR) of a head-and-torso simulator
(HATS) wearing a pair of behind-the-ear hearing aids.
In this study only two-dimensional (2D) sound reproduction was
considered. Although this corresponds to commonly used setups used in
hearing aid research, it brings several limitations: On the perceptual
side, the horizontal plane is considered most important for
localization. For plausible reproduction, however, full immersion is
needed, which implies 3D reproduction.  The technical limitation of 2D
sound reproduction is that a vertical distribution of microphones in
hearing aids can not be tested. Also, for off-center listening, the
spatial distribution of sound intensity differs from the 3D case
\citep{Daniel2001}. Still, in many applications of hearing aid
research a 2D reproduction might be sufficient, because the largest
interaural differences (ILD and ITD) are produced in the horizontal
plane. Additionally, also beamformers mostly operate in this range.

In objective hearing aid algorithm evaluation instrumental performance
measures or performance measures based on perceptual models are
commonly applied \citep{Eneman2008b}. To assess spatial audio
reproduction methods, performance measures of the free field condition
served as a reference in this study. Differences in performance to the
reference condition indicate the lumped effect of the properties of
the reproduction method on algorithm performance. The selection of
performance measures depends on the choice of algorithms to be
tested. Beam patterns \citep[e.g.,][]{Luo2002}, i.e., the
frequency- and azimuth dependent array gains, were analyzed for the
static beamformer. For all algorithms, the signal-to-noise ratio (SNR)
improvement as a function of input SNR and frequency was used. Since
the processed signals are usually presented to a human listener, the
predictions of a perceptual localization model and a monaural
perceptual similarity measure were also applied as baseline measures.

The remainder of the paper is organized as follows: Section
\ref{sec:spatialmethods} describes the used
spatial audio reproduction methods. Algorithm classes are described in
section \ref{sec:hearingaidalgos}, section \ref{sec:performancemeasures} defines
the set of relevant performance measures and the simulation
methods. Results are presented and discussed in section
\ref{sec:results} and \ref{sec:discussion}, respectively. Conclusions
are given in section \ref{sec:conclusions}.

\section{Methods}

\subsection{Spatial audio reproduction methods}\label{sec:spatialmethods}

In this study a spatial audio reproduction method is defined as a set
of driving functions in the form of a set of linear filters ${\bf
  g}_{\bf r}=[g_1,\dots,g_N]$ which generate loudspeaker signals
${\bf x}(t)=[x_1(t),\dots,x_N(t)]$ by convolution of the audio signal
$x_r(t)$ of a single omnidirectional (virtual) sound source at the
position ${\bf r}$ in space with the filters ${\bf g}_{\bf r}$:
\begin{equation}
{\bf x}(t) = {\bf g}_{\bf r} * x_r(t)
\label{eq:drivingsignal}
\end{equation}
A reproduction system is an arrangement of $N$ loudspeakers at the
positions ${\bf s}_k,\,k=1\dots N$. Only regular, horizontal, circular
reproduction systems with even numbers of loudspeakers are addressed
here. Without loss of generality, the center of the reproduction
system is assumed to be at the origin of the coordinate system.

Each driving function $g_k$ can be split into a scalar weight $w_k$
which depends on the position ${\bf r}$ of the source, and a
transmission part $h_k$ which depends only on the distance $||{\bf
  r}||$ of the source from the origin. $h_k$ is the acoustic model of
the source, here consisting of a distance-dependent delay $\tau =
||{\bf r}||/c$ and an attenuation,
\begin{equation}
g_k({\bf r}) = \frac{w_k({\bf r})}{||{\bf r}||}\cdot \delta(\tau),
\end{equation}
where $\delta(\tau)$ is the dirac-function and $c$ is the speed of
sound.  The weights $w_k({\bf r})$ depend on the specific reproduction
method and will be defined below.

\subsubsection{Nearest speaker (NSP)}

The simplest spatial audio reproduction method selects the
loudspeaker $k_{min}$ with the least distance to the
 source for reproduction. The driving weights are thus
\begin{equation}
w_k = \left\{\begin{array}{ll}1 & k=k_{min}\\0 & {\textrm{otherwise}}
  \end{array}\right..
\end{equation}
This reproduction method is equivalent to placing loudspeakers at the
positions of the  sources, which is commonly done in hearing aid
evaluation \citep[e.g.,][]{Greenberg2003}.

\subsubsection{Vector base amplitude panning (VBAP)}

Horizontal vector base amplitude panning as defined by
\citet{Pulkki1997}, uses the closest pair $l,m$ of loudspeakers for
reproduction of a source. Driving weights $w_l$ and $w_m$ are
calculated from the unit vector of the source $\hat{{\bf r}}={{\bf
    r}}\cdot ||{\bf r}||^{-1}$ and the unit vectors of the closest
loudspeakers, $\hat{{\bf s}}_l$ and $\hat{{\bf s}}_m$, with the
loudspeaker matrix ${\bf S}_{l,m}=\left[\hat{{\bf s}}_l\,\hat{{\bf
      s}}_m\right]^{\bf T}$ as
\begin{equation}
\left[w_l\,w_m\right] = \hat{\bf r}^{\bf T}{\bf S}_{l,m}^{-1},
\end{equation}
$w_k=0$ for $k\ne l,m$. With only two loudspeakers, this method
is equivalent to conventional stereo panning.

\subsubsection{Higher order ambisonics (HOA)}

Higher order ambisonics (HOA) is based on the expansion of the sound
field around a single point using spherical harmonics (3D) or
cylindrical harmonics (2D) \citep{Daniel2001}. With increasing
truncation order of the expansion, the size of the area in which the
sound field is well approximated is increasing.
Here, only horizontal higher order ambisonics without near field
compensation is considered. In the case of single virtual sound
sources, i.e., opposed to recorded sources in HOA format, and a
regular reproduction system, the encoding and decoding can be
combined, which drastically reduces complexity \citep{Neukom2007}:
\begin{equation}
w_k = \frac{\sin\left(\frac12 (N-1)\varphi_k\right)}{N \sin\left(\frac12\varphi_k\right)},
\end{equation}
with the azimuth $\varphi_k$ between the source and the $k$th
loudspeaker, and the total number of loudspeakers $N$.  With these
driving weights the method corresponds to the 'basic' HOA method
\citep{Daniel2001}. The minimum number of loudspeakers for a given
ambisonics order $m$ is $N_{min}=2m+1$. However, in this study only
even numbers of $N$ were used, thus the smallest even number of
loudspeakers for a given integer ambisonics order $m$ is
$N_{even}=2(m+1)$. Accordingly, for any given number of loudspeakers
$N$, the largest integer ambisonics order $m=\frac{N}{2}-1$ for even
$N$ was used. Coloration artifacts due to spatial aliasing occur if
the number of loudspeakers is larger than the minimal number for a
given order \citep{Solvang2008}. However, this effect is small in the
current study, because only one more loudspeaker than the minimally
required number of speakers was used.  Spatial aliasing occurs if
\begin{equation}
k r>\frac12 N_{min},\label{eq:aliasing}
\end{equation}
with the wave number $k=\frac{2\pi f}{c}$, the listening position $r$,
i.e., the distance from the origin, and the speed of sound $c$.
This equation can serve as a predictor of the usable bandwidth for a
given number of loudspeakers, e.g., $f\le \frac{c}{4\pi r} N_{min}$,
or as a rough guide for choosing the number of loudspeakers for a
given application and frequency range, $N\ge \frac{4\pi r}{c} f$. In
the case of prediction of binaural listening, $r$ is approximated by
the distance of the ear which is further away from the origin.

\subsubsection{Test signal generation}\label{sec:setup}

The test signals, i.e., the input signals of the hearing aid
processing for the instrumental measures, and the input signals of the
perceptual models, were generated by convolution of the loudspeaker
signals $x_k(t)$ (Eq. \ref{eq:drivingsignal}) with HRIR $h({\bf r},t)$
of a Br\"{u}el \& Kj\ae{}r HATS in an anechoic room
\citep{Kayser2009},
\begin{equation}
x({\bf r},t)=\sum\limits_{k=1}^N
h({\bf s}_k-{\bf r},t) * x_k(t),\label{eq:hrirsim}
\end{equation}
where the star denotes convolution, and ${\bf r}$ is the listener
position. The database provides HRIR for the in-ear microphones of the
HATS as well as the HRIR of six hearing aid microphones, three on each
side. HRIR for a distance $d$ between loudspeaker and the center of
the head of 0.8~m and 3~m exist in the database. In this study the
HRIRs were used for a distance of 3~m, and zero degree elevation,
sampled with a spatial resolution of 5~degrees. For the central
listening position no interpolation of the HRIR was
required. Off-center listening positions, shifted by 0.1 and 0.5~m to
the side, were achieved by applying the distance-dependent gains
$g=3/d$ and delays $\tau=d/c$ to the HRIRs, and by independent
interpolation of the amplitude and phase in the spectrum of the HRIR.
The interpolation method produces amplitude errors below 2~dB and only
negligible errors of group delay when comparing an interpolated HRIR
from two HRIR separated by 10 degree with the corresponding measured
HRIR. In the database the HRIRs are sampled with 5 degrees; thus, the
expected interpolation error is likely to be smaller.
For the experiments based on the perceptual localization model
predictions, the in-ear microphone channels of the HRIR database were
used, corresponding to channels 1 and 2 in \citet{Kayser2009}. For
evaluation of hearing aid algorithms, the appropriate channels for the
respective hearing aid algorithm were used. For the binaural
beamformer these were all six hearing aid channels, for the ADM the
front and rear microphones of the left hearing aid, for the binaural
noise reduction the front microphones of the left and right hearing
aid, and for the single channel noise reduction the front microphone
of the left hearing aid.

\subsubsection{Reference signal generation}\label{sec:refsignalgen}

Reference signals were generated by a convolution of the sound source
signal with the interpolated anechoic HRIR corresponding to the source
direction and distance, which is equivalent to a free field
reproduction of the source signal.

\subsection{Hearing aid algorithms}\label{sec:hearingaidalgos}

Four representative hearing aid algorithms from different classes were
selected for analysis. Three of the algorithms are based on spatially
separated microphones, with different spatial sensitivity. The fourth
algorithm is a standard single-channel noise reduction scheme.  All
algorithms were implemented in C++ within a software platform for
hearing aid algorithm development \citep{Grimm2006}.

%

\subsubsection{Static binaural beamformer}\label{sec:desc_beam}

A binaural multi-microphone beamformer algorithm
\citep{Rohdenburg2007} was selected for the assessment of reproduction
methods because this algorithm, with six spatially separated
microphones, is particularly sensitive to errors in the microphone
signals and the sound field reproduction. From the different versions
of the beamformer introduced by \citet{Rohdenburg2007}, the fixed
minimum variance distortionless response beamformer without a general
sidelobe canceler was chosen. A diffuse noise field was assumed, and a
sampled propagation vector was used, which was matched with the same
HRIR as used in the other parts of this study \citep[see section
  \ref{sec:setup} for details]{Kayser2009}. To preserve binaural cues,
a real-valued time-variant post filter was applied. With this post
filter, the binaural cues of both the target and the noise signal are
preserved.
In a condition with a single target and an artificial diffuse noise,
an absolute SNR improvement of about 6 to 14~dB can be reached, see
Fig.\ \ref{fig:refbenefit}.


\subsubsection{Adaptive differential microphone (ADM)}\label{sec:desc_adm}

The ADM algorithm is based on a front-facing and a back-facing
microphone signal \citep{Elko1995} as typically found in
behind-the-ear hearing aids. These signals are generated by two
delay-and-sum beamformers using a single pair of omnidirectional
microphones. A mixing weight is adapted to minimize the back-facing
signal in the input signal. This algorithm can achieve signal-to-noise
ratio (SNR) improvements of up to 20~dB in anechoic conditions with a
single noise source, and approximately 3 to 6~dB in diffuse-noise
situations, see Fig.\ \ref{fig:refbenefit}.


\subsubsection{Binaural noise reduction}

The binaural noise reduction scheme estimates the interaural coherence
function in multiple frequency bands, to steer a Wiener-like filter
\citep{Kollmeier1993,Wittkop2003}. In this study an omni-directional
variant is used \citep{Grimm2009a,Luts2010}, which estimates the interaural
coherence from the interaural phase difference (IPD) fluctuations. In each
frequency band and time frame, the IPD is measured and transformed onto
the complex plane. The vector strength, i.e., the absolute value of
the low-pass filtered complex-valued IPD is taken as a measure of the
coherence $\gamma$:
\begin{equation}
\gamma = \left|\left\langle e^{i\,\textrm{IPD}}\right\rangle_\tau\right|
\end{equation}
The low pass filter $\langle\cdots\rangle_\tau$ with the time constant
$\tau$ was implemented as a first-order IIR low-pass filter. The
applied gain in each frequency band is $G=\gamma^\beta$. In this
study, the algorithm settings of \citet{Luts2010} were used, i.e.,
$\tau=40$~ms and a frequency-dependent efficiency coefficient $\beta$
ranging from 0 to 0.5.

With this algorithm SNR improvements of about 4 dB can be achieved in
real acoustic environments at frequencies above 1~kHz and at about 0
dB input SNR, see Fig.\ \ref{fig:refbenefit}.


\subsubsection{Single channel noise reduction}

\begin{figure}[htb]
\centering
\includegraphics[width=75mm]{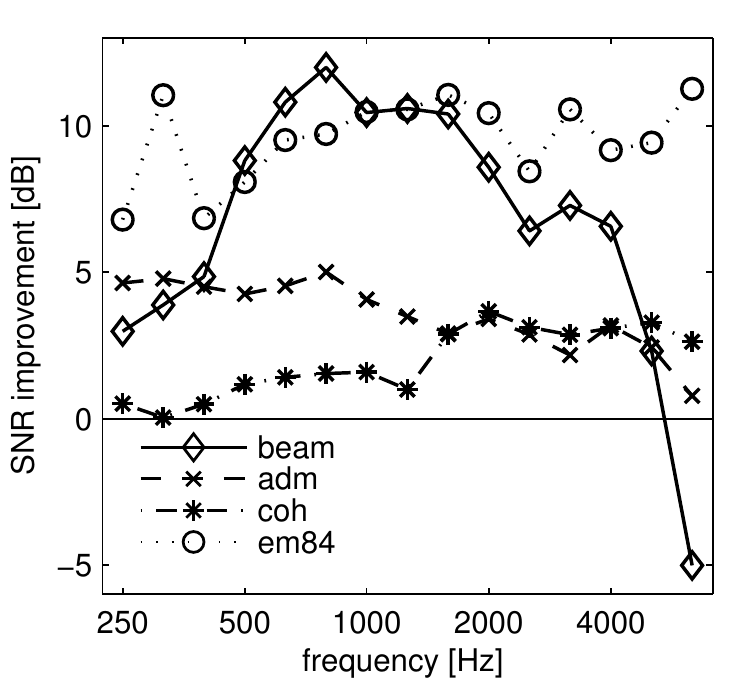}
\caption{Frequency dependent SNR benefit of the four tested algorithms
  in a diffuse noise environment (see Sec. \ref{sec:snranalysis}),
  averaged across all tested input SNRs from $-20$ to 20~dB and
  listening positions, in a free field condition. The SNR benefit is
  an instrumental measure and not necessarily related to any
  perceptual benefit.}\label{fig:refbenefit}
\end{figure}

The single channel noise reduction algorithm after \citet{Ephraim1984}
was chosen as a typical representative of the class of single channel
algorithms. The original algorithm with an optimal noise spectrum
estimator using perfect a-priori knowledge of the noise signal was
used. With this ``oracle'' algorithm an SNR improvement of about 10~dB
was achieved at negative SNRs, independent of the frequency, see
Fig.\ \ref{fig:refbenefit}.


\subsection{Performance measures}\label{sec:performancemeasures}

This study assesses to what extent commonly applied performance
measures are affected by the choice and resolution of the spatial
audio reproduction method. Thus, for each reproduction method and
number of loudspeakers a full technical evaluation of each of the
hearing aid algorithms was performed. The outcome was then compared to
a free field condition as a reference (see
Sec. \ref{sec:refsignalgen}). An error function was defined for each
performance measure to provide a quantitative analysis of differences
compared to the reference.

Suitable measures were applied to each tested algorithm. An analysis
of the beam pattern (Sec. \ref{sec:beamanalysis}) was applied to the
static beamformer. The SNR improvement in a simulated diffuse noise
condition (Sec. \ref{sec:snranalysis}) was applied to all
algorithms. Perceptual localization performance was predicted using a
perceptual localization model, and monaural audio quality was
predicted using a perceptual spectral distance model
(Sec. \ref{sec:percept_model}).

\subsubsection{Beam pattern analysis}\label{sec:beamanalysis}

\begin{figure}[htb]
\centering
\includegraphics[width=75mm]{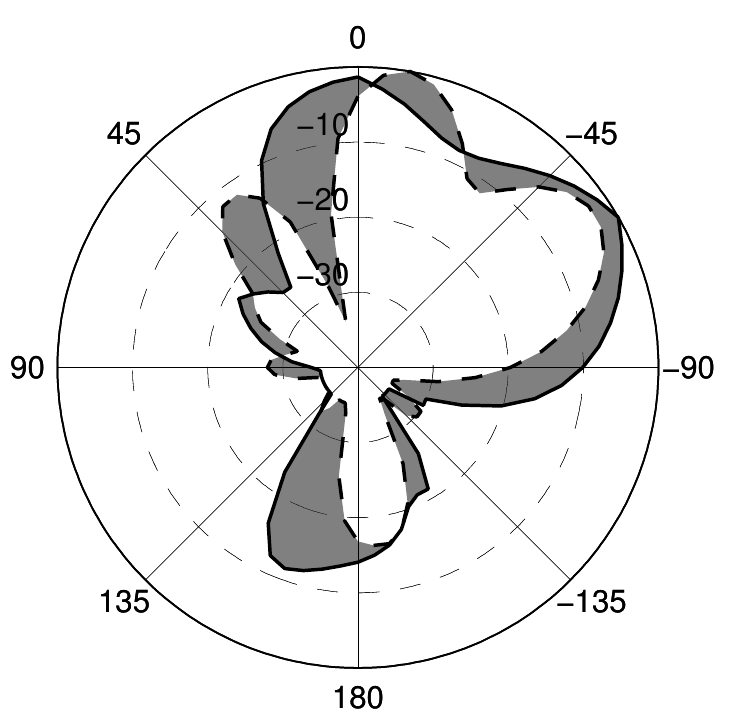}
\caption{Exemplary beam pattern at 2~kHz in the reference condition,
  i.e., free field (dashed line) and with 6-channel VBAP
  (solid line). The RMS of the difference defines the
  beam pattern error (shaded area).}\label{fig:beamerrordef}
\end{figure}

Static beamformers are commonly described by their beam patterns,
i.e., the gain $G(\alpha,f)$ as a function of azimuth $\alpha$ and
frequency $f$. Here, root-mean-square gain deviation of 
$\Delta G = 20\log_{10}(G_{ref})-20\log_{10}(G_{test})$ averaged across all
azimuths $\alpha = 0,5,10,\dots,355$ between the reference beam
pattern $G_{ref}$ (free field) and a test beam pattern
$G_{test}$ (achieved with a specific spatial reproduction method) was
taken as a frequency-dependent measure of reproduction method
performance. The beam pattern was calculated in third-octave
bands. $G_{ref}$ and $G_{test}$ were limited to $-35$~dB for values below $-35$~dB
to avoid an excessive effect of Nulls. The error function can be
written as:
\begin{equation}
E(f) = \sqrt{\sum_\alpha (\Delta G(\alpha,f))^2}\label{eq:beamerror}
\end{equation}
The beam patterns were calculated using HRIRs \citep{Kayser2009} and thus include the
effect of the HATS.  Exemplary beam patterns and a
schematic visualization of the beam error are shown in Figure
\ref{fig:beamerrordef}.

\subsubsection{SNR improvement analysis}\label{sec:snranalysis}

Most hearing aid algorithms modify the SNR to some extent -- some
algorithms like beamformers and noise reduction schemes by intention,
others as an artifact. Thus these algorithms are often characterized
by the SNR improvement behavior, i.e., the difference of the SNR at
the output $R_o$ and at the input $R_i$ as a function of input SNR.

\begin{figure}[htb]
\centering
\includegraphics[width=75mm]{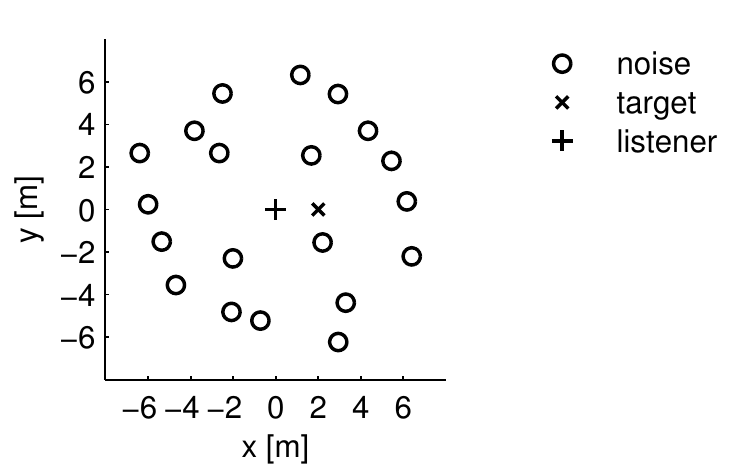}
\caption{Simulated diffuse noise situation for the SNR evaluation. Circles
  indicate the positions of noise sources, each radiating a
  cafeteria-like noise signal. The diagonal cross indicates the
  position of the frontal target speech
  signal.}\label{fig:snrcafeteria}
\end{figure}

\begin{figure}[htb]
\centering
\includegraphics[width=75mm]{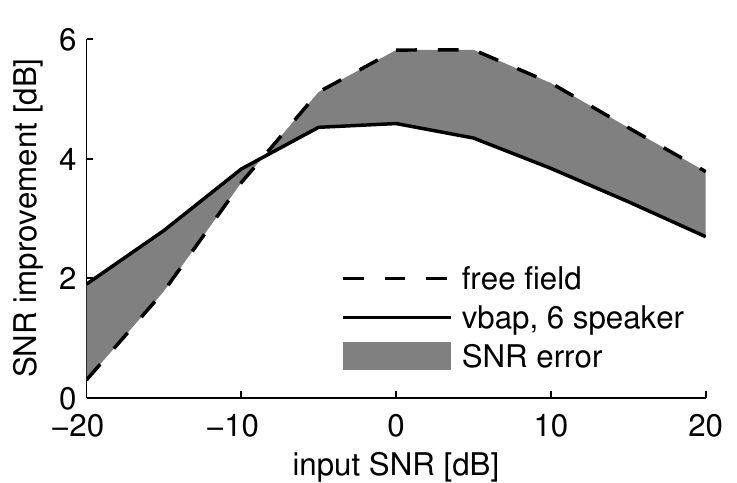}
\caption{Exemplary SNR improvement of binaural noise reduction in the
  reference condition, i.e., free field (dashed line) and
  with 6-channel VBAP (solid line). The RMS of the difference averaged
  across all input SNRs defines the SNR improvement error (shaded
  area).}\label{fig:snrerrordef}
\end{figure}

Here, the SNR behavior in a diffuse noise situation with a single
target speech signal from the front and 20 spatially distributed
cafeteria-noise sources (see Figure \ref{fig:snrcafeteria}) was chosen
as a measure of reproduction method performance. The target stimulus
was a 8.4-second segment of a female monologue.  The diffuse
noise environment was created by adding cafeteria-noise sound sources
from different directions and with an attenuation corresponding to
the respective distances. Each of the sources was simulated using the method
described in \ref{sec:setup}. Early reflections and diffuse
reverberation were not added. The noise stimuli were non-overlapping
segments taken from a single-channel recording in a real cafeteria, containing a
clutter of cutlery noises, babble and moved chairs. The long-term SNR
improvement of the hearing aid algorithm $\Delta R(f)$ was estimated
in third-octave bands, for nine different nominal broad-band input
SNRs $R_{i,n} = -20,-15,-10,\dots,20$~dB. As error measure the
root-mean-square difference between reference condition with free
field and the test condition with application of the spatial audio
reproduction method was computed:
\begin{equation}
E(f) = \sqrt{\frac19 \sum_{R_{i,n}}\left( \Delta R_{ref}(f) -  \Delta R_{test}(f)\right)^2}
\label{eq:snrerror}
\end{equation}
An exemplary SNR improvement of the binaural noise reduction algorithm
is shown in Figure \ref{fig:snrerrordef}.

\subsubsection{Analysis of errors in perceptual measures of localization and spectral distortion}\label{sec:percept_model}

For modeling source localization, the binaural model of
\citet{Dietz2011} was used.  It estimates the interaural time
difference (ITD) and the interaural level difference (ILD) in auditory
frequency bands.  An interaural coherence function $\gamma$ is
calculated to select only those ``glimpses'', i.e., time-frequency
signal components, with a high interaural coherence ($\gamma > 0.98$)
and a rising coherence slope. Only these time-frequency components are
assumed to contain reliable perceptual binaural cues.  In those
frequency bands where temporal fine-structure is available to human
listeners (12 bands with center frequencies from 236 to 1296~Hz), the
fine-structure ITD is used, and ambiguities are resolved by means of
the sign of the ILD, i.e., ILD is only used for disambiguation and not
explicitly for estimation of source direction. Direction of arrival
(DOA) estimates are interpolated from a look-up table, derived from
anechoic HRIRs of the HATS.  In this model version, envelope ITDs are
used in frequency bands from 1296~Hz to 4~kHz to gather a second DOA
estimate.  Since only single sources were taken into account in this
study, the estimated direction of arrival $\alpha$ was averaged across
all selected glimpses of the test stimulus, for fine-structure
frequency bands and envelope frequency bands separately.

Based on the localization model output, the perceptual localization
error (PLE) was defined as the RMS difference between the estimated
direction of arrival in a free field, $\alpha_{ref}$, and the
estimated direction of arrival measured with the tested reproduction
method, $\alpha_{test}$, for all nominal target directions from $-75$
to 75 degrees in steps of 5 degrees. This limited azimuth range was
chosen to avoid problems of the model at lateral signal sources. A
small PLE indicates that the predicted perceptual localization does
not depend on the reproduction method, whereas for high PLE values the
reproduction method has an effect on the predicted perceptual
localization. Comparison with the reference condition (free field) has
the advantage of separating out the effect of the reproduction methods
from the perceptual localization performance as modeled by the
binaural model. This means that, similar to the more technical
measures introduced above, this measure does not rate the absolute
perceptual localization performance in the tested methods.

Monaural perceptual features were assessed by a model for predicting
the perceived naturalness of sounds subjected to spectral distortion
\citep{Moore2004}. The spectral distance between two stimuli (free
field and reproduced sound field, in this case) is calculated by a
comparison of excitation patterns created by an auditory filter
bank. Absolute differences, differences in ripple and spectral slope
are combined by a weighted sum to form a scalar spectral distance
measure. In the original paper, the spectral distance was further
transformed into a prediction of perceived naturalness; here, the
distance measure was directly used. In contrast to the binaural model,
this measure rates only monaural spectral features, e.g., changes in
coloration.

Since the spectral distance represents already a difference between
signals, it was averaged across target azimuths from $-75$ to 75
degrees in steps of 5 degrees. As a reference signal the free field
condition was used.  

The stimulus used for all perceptual evaluations was a 8.4-second
segment of a female monologue.

\subsubsection{Error criteria}

The theoretical spatial aliasing criterion (Eq. \ref{eq:aliasing}) can
be used as an estimate of the usable frequency range for a given
number of loudspeakers and size of the listening area. Likewise it can
be used as an estimate of the minimal number of loudspeakers for a
given frequency range and listening area, or as a predictor of usable
listening area for a given number of loudspeakers and frequency
range. However, it does not characterize the interaction between the
hearing aid algorithm and the reproduction method. Therefore, for each
instrumental measure, an error criterion is desirable providing an
algorithm-specific guide for the selection of the appropriate
reproduction method and number of loudspeakers, or as a predictor of
the usable frequency range and listening area size for a given
algorithm and number of loudspeakers. Since the instrumental measures
are not directly related to perception, the choice of the threshold is
somewhat arbitrary. To allow for a comparison across reproduction
methods and algorithms, the threshold was chosen so as to best
approximate Eq.\ \ref{eq:aliasing} in a reference condition. As
reference condition the 10~cm off-center listening position with HOA
reproduction was used, because Eq.\ \ref{eq:aliasing} is valid for
HOA.  In particular, the threshold criterion was set such that for each
measure the number of data points meeting the respective criterion was
the same as for the theoretical threshold criterion in the reference
condition.  The resulting error criteria are given in Table
\ref{tab:criterion}. In case of the SNR measure, the criterion
corresponds to roughly 5\% of the maximum algorithm-specific benefit.

\begin{table*}
\begin{tabular}{llc}
\hline
Measure    & algorithm                        & criterion \\
\hline
beam error & binaural beamformer              & 5.7~dB    \\
SNR error  & binaural beamformer              & 0.75~dB   \\
SNR error  & adaptive differential microphone & 0.42~dB   \\
SNR error  & binaural noise reduction         & 0.42~dB   \\
SNR error  & single channel noise reduction   & 0.65~dB   \\
\hline
\end{tabular}
\caption{Error criteria used in the instrumental measures for
  comparison with the spatial aliasing criterion
  (Eq. \ref{eq:aliasing}).}\label{tab:criterion}
\end{table*}

\subsection{Evaluated parameter space}

All reproduction methods were evaluated with $N=$ 4, 6, 8, 12, 18, 24,
36 and 72 loudspeakers, resulting in an angular distance between
loudspeakers of 90, 60, 45, 30, 20, 15, 10 and 5 degrees, and a
spatial distance of 2.83, 2.00, 1.53, 1.04, 0.69, 0.52, 0.35 and
0.17~m, respectively. Three listening positions were evaluated, one in
the origin, one 0.1~m to the side, corresponding to the range of head
movements of a seated listener, and one 0.5~m to the side,
corresponding to the range of torso movements of a listener.

\section{Results}\label{sec:results}

\subsection{Beam pattern error}\label{sec:res_beampattern}

\begin{figure}[htbp]
\centering
\includegraphics[width=75mm]{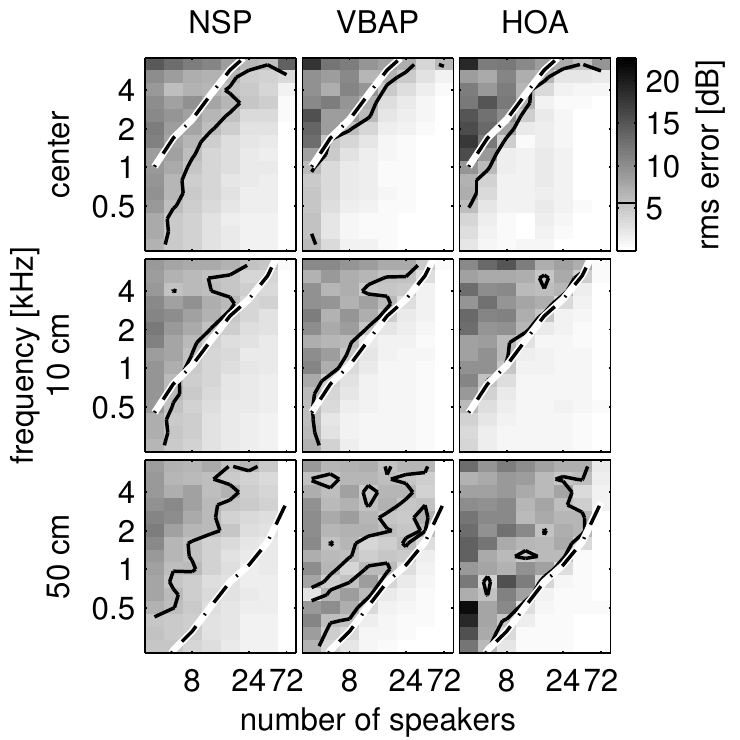}
\caption{Beam pattern error as a function of frequency and number of
  loudspeakers, with a contour line for the 5.7~dB beam error
  criterion (solid line). Additionally, the HOA aliasing criterion
  (Eq. \ref{eq:aliasing}) for the ear with the largest distance to the
  center is marked by a dashed-dotted line. In the three top panels
  the listener was positioned in the center of the listening area. In
  the middle panels, the listener was moved 10~cm to the left side
  (corresponding to head movements), and in the bottom panels the
  listener was moved 0.5~m to the left side (corresponding to torso
  movements).}\label{fig:beamerror}
\end{figure}

The beam error (Eq. \ref{eq:beamerror}, Sec. \ref{sec:beamanalysis})
of the static binaural beamformer as a function of the number of
loudspeakers and frequency is shown in Figure \ref{fig:beamerror}. For
all reproduction methods the usable bandwidth is increasing with the
number of loudspeakers, and decreasing with the distance of the
listener position from the origin. For NSP the beam error is caused by
the sub-sampling of the beam pattern, and is largely independent from
the listening position. For VBAP and HOA the beam error depends on the
listening position. For off-center listening position the beam error
criterion is well approximated by the spatial aliasing criterion
(Eq. \ref{eq:aliasing}). VBAP and HOA show essentially the same
behavior except for very low number of loudspeakers, where HOA
performs slightly better.
For example, to achieve a bandwidth of 2~kHz in the central listening
position, 24 loudspeakers would be required for NSP, and 12 for VBAP
and HOA. If for example 24 loudspeakers are available, the usable
bandwidth in the central listening position is 2~kHz for NSP, 4~kHz for
VBAP, and 5~kHz for HOA; in the 50~cm off-center listening position
the same number of loudspeakers would lead to a usable bandwidth of
4~kHz with NSP, and 1~kHz with VBAP and HOA.


\begin{figure}[htbp]
\centering
\includegraphics[width=75mm]{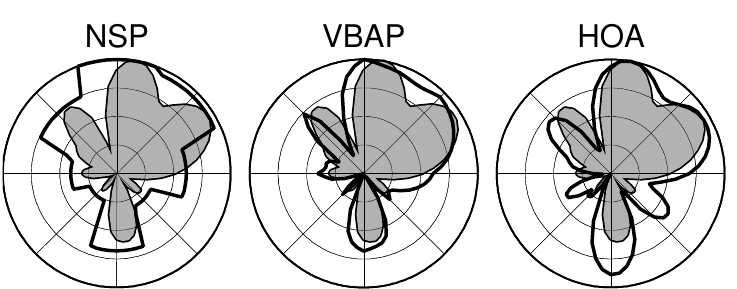}
\caption{Exemplary beam pattern of the binaural beamformer, measured
  at 2~kHz with 8 loudspeakers using NSP (left panel), VBAP (center
  panel) and HOA (right panel). The shaded area denotes the beam
  pattern in the reference condition (free
  field).}\label{fig:beampatternex}
\end{figure}

The exemplary beam patterns (Fig. \ref{fig:beampatternex}) illustrate
the differences in spatial interpolation between the three spatial
reproduction methods. The effect of nearest neighbor sampling in the
NSP method is obvious. However, VBAP interpolates only between two
sources, which results in noncontinuous derivatives over the azimuth
(e.g., sharp tips of the side lobes). With HOA the gain is
continuously differentiable.

\subsection{SNR behavior of hearing aid algorithms}

\begin{figure}[htbp]
\centering
\includegraphics[width=75mm]{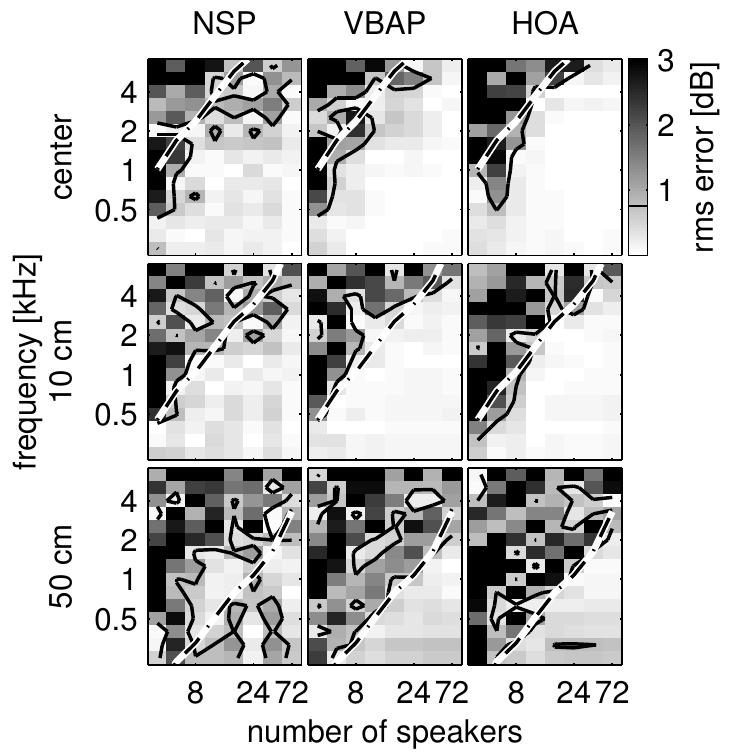}
\caption{SNR error of the binaural beamformer as a function of
  frequency and number of loudspeakers, with a contour line for the SNR
  error criterion of 0.75~dB (solid line). In the three top panels the
  listener was positioned in the center of the listening area. In the
  middle panels, the listener was moved 10~cm to the left side
  (corresponding to head movements), and in the bottom panels the
  listener was moved 0.5~m to the left side (corresponding to torso
  movements).}\label{fig:snrerrorbeam}
\end{figure}

\begin{figure}[htbp]
\centering
\includegraphics[width=75mm]{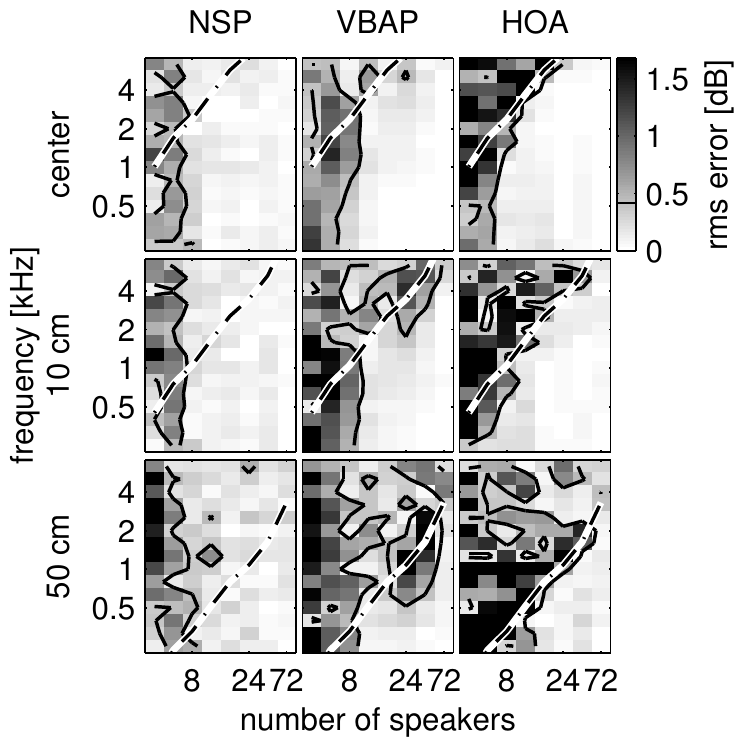}
\caption{Same as Figure \ref{fig:snrerrorbeam}, but for the
  adaptive differential microphone, with an SNR error criterion of
  0.42~dB.}\label{fig:snrerroradm}
\end{figure}

\begin{figure}[htbp]
\centering
\includegraphics[width=75mm]{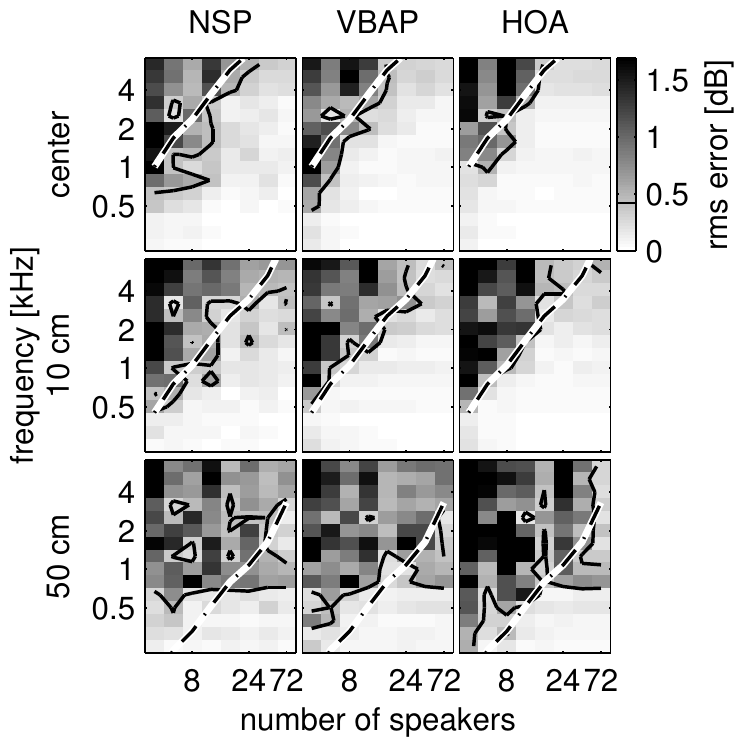}
\caption{Same as Figure \ref{fig:snrerrorbeam}, but for the
  binaural noise reduction, with an SNR error criterion of
  0.42~dB.}\label{fig:snrerrorcoh}
\end{figure}

\begin{figure}[htbp]
\centering
\includegraphics[width=75mm]{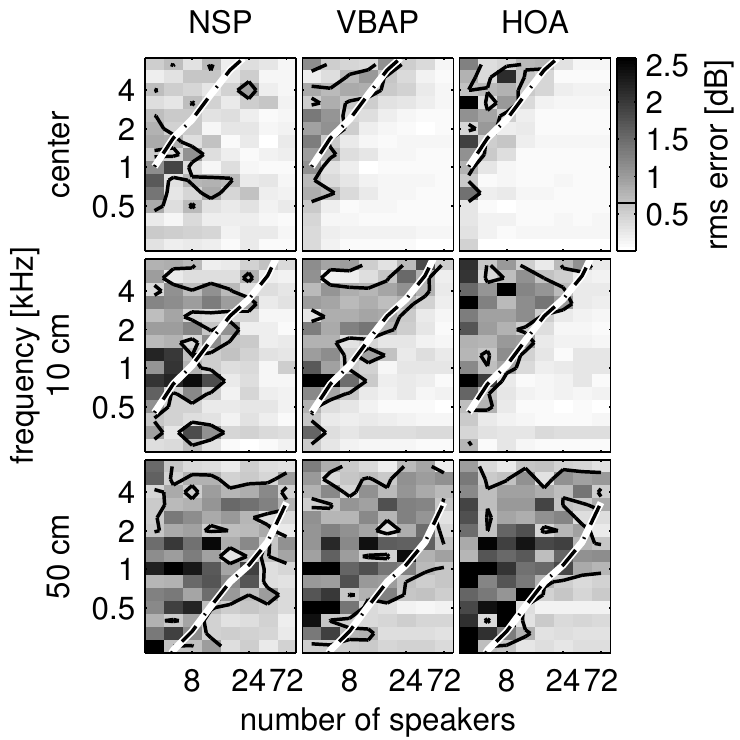}
\caption{Same as Figure \ref{fig:snrerrorbeam}, but for the single
  channel noise reduction, with an SNR error criterion of
  0.65~dB.}\label{fig:snrerrorem84}
\end{figure}

The SNR error (Eq. \ref{eq:snrerror}) of the four tested hearing aid
algorithms in a simulated diffuse noise environment with 20 noise sources
and a frontal target source is shown in Figures \ref{fig:snrerrorbeam}
to \ref{fig:snrerrorem84}.

The SNR error of the binaural beamformer, Fig. \ref{fig:snrerrorbeam},
decreases with increasing number of loudspeakers and with decreasing
frequency, similar to the beam error. The SNR error criterion of
0.75~dB is well predicted by the theoretical aliasing criterion
Eq. \ref{eq:aliasing}, for VBAP and HOA. With NSP the SNR error
criterion can only be reached up to 2~kHz even for large numbers of
loudspeakers. In the 50~cm off-center listening position the SNR error
is above the threshold also for low frequencies.
For example, if 24 loudspeakers are available, the usable bandwidth 
in the central listening position is 2~kHz for NSP, 4~kHz for VBAP, and 5~kHz
for HOA; in the 50~cm off-center listening position the same number of
loudspeakers would lead to a usable bandwidth of 500~Hz with NSP, and
1~kHz with VBAP and HOA.

The adaptive differential microphone, Fig. \ref{fig:snrerroradm},
shows a completely different SNR behavior: The SNR error is smallest
for NSP; here it is below the threshold criterion of 0.3~dB at most
frequencies and all listening positions as soon as more than 8
loudspeakers are used for reproduction. With HOA the SNR error
criterion is similar to the spatial aliasing criterion of
Eq. \ref{eq:aliasing}, limiting the usable bandwidth for low numbers
of loudspeaker. With VBAP the performance is between NSP and HOA.

The binaural noise reduction algorithm, Fig. \ref{fig:snrerrorcoh},
draws again a different picture: Here, the SNR error is more or less
independent from the reproduction method. For the central listening
position and the 10~cm off-center listening position the spatial
aliasing criterion, Eq. \ref{eq:aliasing}, predicts the performance
well for all reproduction methods. At low frequencies the SNR error is
low in all conditions. This is caused by the fact that the algorithm
has no significant effect on low frequency components, because the
interaural coherence is always high.
For example, to achieve a bandwidth of 2~kHz in the central listening
position, 18 loudspeakers would be required for NSP and VBAP, and 12
for HOA. If for example 24 loudspeakers are available, the usable
bandwidth is in central listening position 4~kHz for NSP and VBAP, and
6~kHz for HOA.

The SNR error of the single channel noise reduction, shown in Fig
\ref{fig:snrerrorem84}, shows a similar behavior as the binaural noise
reduction scheme, even without any explicit spatial sensitivity. For
VBAP and HOA the SNR error criterion is again approximated by the
aliasing criterion. As a tendency also the SNR error criterion with
NSP is predicted by the aliasing criterion. However, the SNR error
shows now more frequency dependency than for the other tested
algorithms, i.e., the effect of the number of loudspeakers used in the
reproduction is smaller. The selection of reproduction method has no
clear effect on the SNR error.

\subsection{Perceptual model predictions}\label{sec:data_perceptual}

\begin{figure}[htbp]
\centering
\includegraphics[width=75mm]{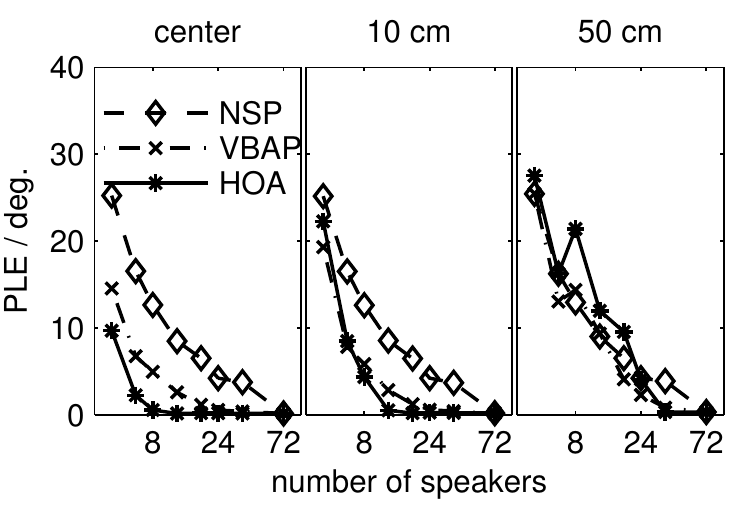}
\caption{Perceptual localization error (PLE), as predicted by the
  perceptual binaural localization model of \citet{Dietz2011}. The
  left panel shows data for the central listening position, the center
  and right panels for the listening position 0.1~m and 0.5~m to the
  left of the center, respectively. For NSP the PLE is half of the
  angular distance between the loudspeakers.}\label{fig:perc_doa}
\end{figure}

\begin{figure}[htbp]
\centering
\includegraphics[width=75mm]{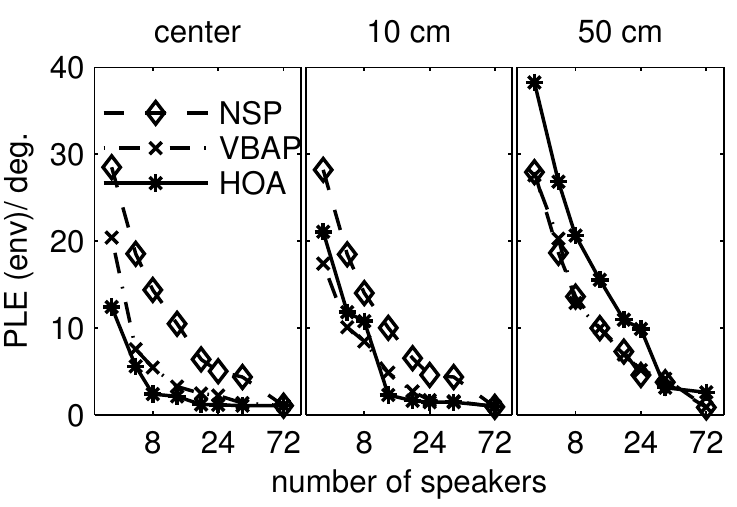}
\caption{Same as Figure \ref{fig:perc_doa}, but for the direction of arrival estimation based on the envelope in frequency bands above 1.3~kHz.}\label{fig:perc_doa_env}
\end{figure}

The perceptual localization error (PLE) in the three listening
positions is shown in Figure \ref{fig:perc_doa}. For NSP the PLE is
half of the angular distance between the loudspeakers. In the central
listening position HOA reproduces the DOA the best, with a negligible
PLE starting with 8 loudspeakers. For off-center listening positions
the PLE of HOA and VBAP increases, and is the same as for NSP in the
50~cm off-center listening position. The same ranking of errors can be
observed when estimating the direction of arrival based on the
envelope ITD in frequency bands above 1.3~kHz, see Figure
\ref{fig:perc_doa_env}.

\begin{figure}[htbp]
\centering
\includegraphics[width=75mm]{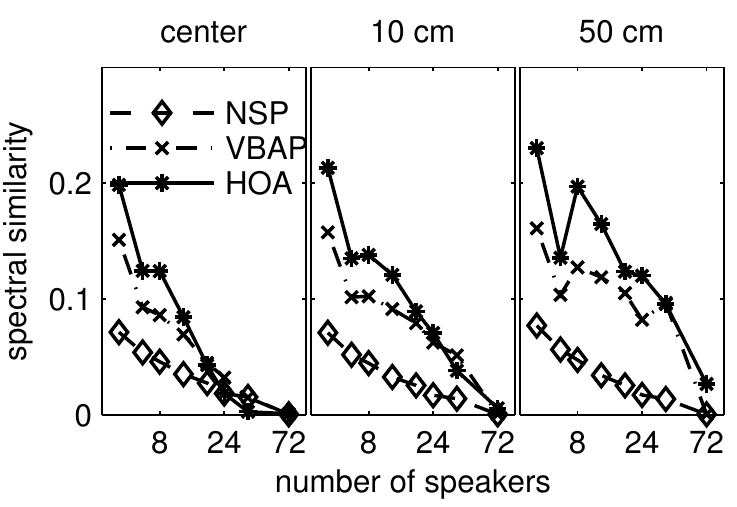}
\caption{ Monaural spectral distance, as predicted by the perceptual
  model of \citet{Moore2004}. Values of 0 correspond to no perceptual
  difference to the reference free-field condition. The data were
  averaged across all tested target directions from $-75$ to 75
  degrees. Except for the central listening position with at least 18
  loudspeakers, HOA produces the largest monaural quality
  degradation. }\label{fig:perc_psm}
\end{figure}

The perceptual spectral distance between the virtual sound source
(reference) and the reproduced source is shown in
Fig.\ \ref{fig:perc_psm}. The test stimulus was the same speech signal
as in the SNR measurements, see \ref{sec:snranalysis} for details.
For NSP the distance is determined only by timbral changes caused by
the spatial sampling of the HRIRs. For the other reproduction methods,
also spectral changes caused by spatial aliasing contribute to an
increased spectral distance.

All of the spectral distance values are below 0.25. A comparison with
the subjective data provided by \citet{Moore2004} indicates that even
the largest spectral distance measured in this study corresponds to
the highest rating of naturalness for speech.

\section{Discussion}\label{sec:discussion}

When loudspeaker-based spatial audio reproduction methods are involved
in the evaluation of hearing aids, the question arises to what extend
the results are influenced by the reproduction methods.  All
reproduction methods have a physical limitation caused by the spatial
aliasing, which means that above a certain frequency the reproduced
signals contain spectral and spatial artifacts. The reproduction
methods evaluated in this study represent a trade-off between minimal
spectral artifacts and maximal spatial limitations (NSP) at one end,
and maximal spectral artifacts combined with minimal spatial artifacts
(HOA) on the other end of the scale. However, the picture is not that
clear for hearing aid evaluations: Depending on the microphone
positions and signal processing of the hearing aids, spatial
artifacts can translate into spectral ones and vice versa. The SNR
error of the single channel noise reduction
(Fig. \ref{fig:snrerrorem84}) may serve as an example: Although the
algorithm uses only a single microphone, the error of the SNR
performance measure follows more or less the spatial aliasing relation
with all reproduction methods. Since the only input parameter to the
algorithm is the frequency dependent short-time SNR, it can be
concluded that spatial resolution can translate into spectral artifacts
as soon as multiple virtual sources are involved.

It is well known from the literature that broadband signal to noise
ratio is not directly related to speech intelligibility, e.g., single
channel noise reduction often increases the SNR without any positive
effect on speech intelligibility. Here, the broadband SNR is used as a
differential algorithm performance measure. For a prediction of speech
intelligibility, however, the SNR remains a major component. Commonly
used models for speech intelligibility prediction are segmental SNR
\citep{Mermelstein1979}, frequency weighted SNR (SII)
\citep{ANSI-S3.51997}, or some form of signal-to-noise ratio after
modeling the auditory periphery \citep{Christiansen2010}. The
application of a much simpler broadband measure seems applicable in
the context of this study, because not the absolute performance of the
algorithms and their user benefit was of interest, but rather the
validation of multichannel loudspeaker reproduction. If the tested
reproduction methods are unable to reproduce the effect of algorithms
on broadband SNR, it is likely that realistic speech intelligibility
measurements would not be warranted either.

The threshold criteria for the instrumental measures of this study
were derived from the distribution of the measured data, to allow a
comparison with the theoretical spatial aliasing criterion. For the
SNR-based measure, the resulting criteria are related to the maximal
algorithm SNR benefit. For all algorithms, the threshold is in the
range between 0.42 and 0.75~dB, which is comparable to the resolution
of speech reception threshold measurements.

Multichannel loudspeaker audio reproduction methods may introduce
several different artifacts. This study focuses only on artifacts
which explicitly relate to hearing aid algorithm performance. This
does not imply that other artifacts, such as reduced perceptual
spatial resolution, increase of apparent source width, or head
movement related time-varying coloration don't have an effect on
naturalness and immersion when listening with hearing aids. However,
these classes of artifacts have been thoroughly described in the
literature
\citep[e.g.,][]{Landone1999,Daniel2001,Daniel2003,Carlsson2004,Pulkki2005,Ahrens2008,Benjamin2010,Bertet2013,Heeren2014},
and an in-depth analysis of these artifacts would go beyond the scope
of this study.

For the purpose of this study, it was necessary to quantify any
changes in technical algorithm performance induced by sound field
approximations relative to the free sound field in a systematic,
significant and sensitive way. It is therefore not claimed nor
necessary that the applied instrumental measures reflect subjective
performance, they only need to be sensitive to relative changes in
algorithm performance induced by small changes of the sound field, in
particular for speech sounds. The broadband SNR appears suitable for
this purpose, as it is an established measure, integrates in a
meaningful way across frequency and is robust against systematic small
changes of the absolute transfer characteristics.

Different hearing aid algorithms are designed to provide benefit in
different acoustic environments. Here, all algorithms were tested in
only one diffuse noise environment. In other environments, the
absolute benefit of algorithms might be different, and they might be
more sensitive to the spatial resolution of the reproduction system;
it remains to be studied to what extent the results in the diffuse
noise environment may be generalized. The results indicate that
besides some algorithm specific differences, the theoretic spatial
aliasing criterion is a good first estimate for the effect on
reproduction method performance. This suggests that the selection of
the test environment was reasonable, and results may extent to other
environments. Additionally, in the diffuse noise environment all of
the tested algorithms are expected to provide some benefit, whereas in
other environments, e.g., a single target with a single noise source,
some algorithms are known to fail.

In this study only broadband spatial audio reproduction methods were
tested. By the use of optimal reproduction methods in different
frequency ranges the perceptual artifacts can be reduced at high
frequencies, especially for off-center listening positions. This is
common practice in HOA applications, where often 'basic' decoding is
applied at low frequencies, and 'max-rE' decoding at higher
frequencies \citep{Daniel2001}.

Here, instrumental performance measures were applied to the assessment
of 2D audio reproduction. For plausibility of virtual environments,
however, the technical precision of reproduction might be less
important than a full immersion, as it would only be achieved by 3D
audio reproduction. In hearing-aid research, however, most established
evaluation procedures employ high-resolution 2D spatial setups and
current algorithms mainly consider horizontal spatial properties, so
that a compromise solution may be a mixed system with high horizontal
resolution for sources which require high spatial resolution, and a
low resolution 3D system primarily for immersion
\citep[e.g.,][]{Grimm2013,Grimm2014a}.

This study is based on simulations in a free sound field, as would be
achievable by placing the loudspeaker array in an anechoic room. For
practical applications, however, most systems would be located in
regular rooms, optimally with some sound absorbing acoustic
treatment. Accordingly it is of interest to know to what extent the
results of the current study may be transferred to such real
rooms. Obviously rooms with salient room resonances may create
standing waves, which will reduce localization performance for any
reproduction method. Also early lateral reflections and large amount
of reverberation decrease localization performance, as was shown by
\citet{Hartmann1983}. On the other hand, the monaural artifact of
spectral coloration due to comb filter artifacts introduced by the
VBAP and HOA reproduction methods in off-center listening positions,
which is clearly perceivable in anechoic conditions when the listener
is moving laterally, can be substantially masked by a moderate amount
of room reverberation. All together, the main differences between the
analyzed reproduction systems therefore will remain in real
rooms. Appropriate acoustic treatment is recommended, particularly
when physically correct reproduction is of importance.

\section{Conclusions}\label{sec:conclusions}

All tested spatial reproduction methods are suitable for the
assessment of hearing aid algorithm performance. However, the optimal
system and its required number of loudspeakers depend on the type of
hearing aid algorithm as well as on bandwidth requirements.

In tasks which require a high spatial resolution, such as an analysis
of beam patterns of directional algorithms, higher order ambisonics
and vector base amplitude panning performed best.

In tasks which analyze the SNR behavior of hearing aid algorithms the
optimal reproduction method depends on the algorithm class: The
performance of binaural noise reduction is largely independent of the
reproduction method, and depends only on the number of loudspeakers
and the listening position. The analysis of an adaptive differential
microphone revealed that the theoretical free-field SNR behavior is
best reproduced with the selection of the nearest speaker for each
source. In that case the performance does not depend on the number of
loudspeakers or listening position, if at least eight loudspeakers are
used.

The theoretical free-field SNR behavior of a binaural beamformer is
best reproduced in the central listening position by higher order
ambisonics. Also perceptual localization performance in the central
listening position is best reproduced by higher order ambisonics --
here the deviation from free field simulation is negligible even with
only eight loudspeakers. However, for off-center listening positions
the advantage of higher order ambisonics vanishes.

The data also show that care has to be taken in selecting the
appropriate reproduction method even when only algorithms are involved
that do not explicitly depend on spatial sound field properties, such
as single channel noise reduction. Furthermore, it can be concluded
that even the selection of discrete speakers, which is free of spatial
aliasing for a single source, can lead to typical spatial aliasing
artifacts when multiple sources are reproduced.

As a rough guideline the data can be summarized as follows:
\begin{itemize}
\item With fewer than eight loudspeakers, the performance measure
  criteria are not matched for most tested conditions.
\item For a beam pattern analysis and 4~kHz bandwidth, 18 loudspeakers
  are required in the central listening position (no head movements),
  36 loudspeakers are required in 10~cm off-center listening position
  (head movements allowed), and 72 loudspeakers are required in the
  50~cm off-center listening position (head- and torso movements
  allowed). For a beam pattern analysis, VBAP and HOA appear to be the
  best choice.
\item The SNR behavior of the adaptive differential microphone (ADM)
  in complex acoustic scenarios is best reproduced with discrete
  speakers (NSP) in all listening positions. Using more than 8
  loudspeakers does not provide any benefit in this condition.
\item The SNR behavior of single channel noise reduction is best
  reproduced using VBAP or HOA. This indicates that spatial audio
  reproduction methods which interpolate between loudspeakers can be
  beneficial even for hearing aid algorithms which do not explicitly
  depend on spatial properties of the sound field.
\end{itemize}

\section*{Acknowledgment}

This study was funded by the German Research Foundation DFG, research
unit 1732 (``Individualisierte H{\"o}rakustik'').

\bibliographystyle{abbrvnat}
\bibliography{grimm_multichannel}

\end{document}